# Ion heating dynamics in solid buried layer targets irradiated by ultra-short intense laser pulses

L. G. Huang, [1,2,3,a)] M. Bussmann, [2] T. Kluge, [2] A. L. Lei, [5] W. Yu, [1] and T. E. Cowan [2,4]

[1] *Shanghai Institute of Optics and Fine Mechanics, Chinese Academy of Sciences, 201800 Shanghai, China*

[2] *Helmholtz-Zentrum Dresden-Rossendorf, 01328 Dresden, Germany*

[3] *University of Chinese Academy of Sciences, 100049 Beijing, China*

[4] *Technische Universität Dresden, 01062 Dresden, Germany*

[5] *Shanghai Institute of Laser Plasma, 201800 Shanghai, China*

We investigate bulk ion heating in solid buried layer targets irradiated by ultra-short laser pulses of relativistic intensities using particle-in-cell simulations. Our study focuses on a $CD_2$-Al-$CD_2$ sandwich target geometry. We find enhanced deuteron ion heating in a layer compressed by the expanding aluminium layer. A pressure gradient created at the Al-$CD_2$ interface pushes this layer of deuteron ions towards the outer regions of the target. During its passage through the target, deuteron ions are constantly injected into this layer. Our simulations suggest that the directed collective outward motion of the layer is converted into thermal motion inside the layer, leading to deuteron temperatures higher than those found in the rest of the target. This enhanced heating can already be observed at laser pulse durations as low as 100 femtoseconds. Thus, detailed experimental surveys at repetition rates of several ten laser shots per minute are in reach at current high-power laser systems, which would allow for probing and optimizing the heating dynamics.

a) Electronic mail: lingen.huang@hzdr.de

# I. INTRODUCTION

Bulk ion heating driven by intense lasers is of great interest as it provides experimental access to the ultra-fast dynamics in dense plasmas as they occur in stellar shocks or inertial fusion[1-6]. As of yet, most experimental studies focus on the long-term evolution of ion heating on the order of several hundred picoseconds to nanoseconds[7-10]. This is due to the fact that heating a significant part of all ions in a solid-density target to temperatures of several hundred eV is usually only possible using high-energy laser pulses of several hundred Joule and picosecond to nanosecond time scale. With currently existing laser systems such laser pulses can only be delivered at repetition rates on the scale of few shots per hour, preventing detailed studies of the heating dynamics. Consequently, many simulation studies neglect the particle dynamics in the heating process happening on the sub-picosecond scale and focus on the long-term temperature evolution in the laser-driven plasma[11-13].

The kinetic simulations presented in this work provide an exhaustive parameter study of bulk ion heating in solid $CD_2$-Al-$CD_2$ buried layer targets irradiated by ultra-short intense laser pulses with intensities ranging from $2\times10^{19}\,\text{W}/\text{cm}^2$ to $5\times10^{20}\,\text{W}/\text{cm}^2$ and pulse durations ranging from $100\,\text{fs}$ to $500\,\text{fs}$. Laser-driven generation of hot electrons at the target front side as well as electron-electron, electron-ion and ion-ion interactions within the target bulk are resolved on a sub-femtosecond scale. We are therefore able to connect the time scale of the ultra-short laser target interaction at the front side to the bulk ion heating time scale. Our results indicate that with buried layer targets these time scales can be efficiently decoupled and ultra-short, high-intensity lasers with repetition rates of few shots per second can be used to study bulk ion heating. We find compression of the plastic layer by a factor of approximately 1.5, from $1.1\,\text{g}/\text{cm}^3$ to $\sim 1.65\,\text{g}/\text{cm}^3$, and bulk deuteron ion temperatures up to $800\,\text{eV}$ within the compressed layer. Our analysis shows that the rate of fractional energy transfer from the directed motion of deuterons into thermal motion during the expansion of the compressed layer decreases from $0.7 \text{ ps}^{-1}$ to $0.2 \text{ ps}^{-1}$ with increasing laser intensity. We however find that this decrease is more than compensated by the increase in deuteron directed velocities observed at higher laser intensities, leading to bulk deuteron temperatures of several hundred eV even for the case of ultra-short laser pulse durations.

## II. METHODS

### A. Simulation parameters

All simulations presented in this work have been performed using the 2D3V particle-in-cell code iPICLS2D[14] and include electron-electron, electron-ion and ion-ion collisions as well as collisional and field ionization. In our simulations, the solid buried layer target with $26 \text{ μm}$ height consists of one Al layer with $1 \text{μm}$ thickness coated by two $CD_2$ plastic layers with $2 \text{ μm}$ thickness each as can be seen in Fig. 1(a). The front surface of the target is positioned at $27.5 \text{ μm}$ distance from the left border of the simulation box. The target is centred vertically with respect to the laser axis. The laser pulse coming from the left side of the simulation box irradiates the target at normal incidence with wavelength $\lambda_0 = 1.054 \text{ μm}$. It is modelled using a spatial and temporal Gaussian envelope with peak laser intensities $I_0$ varying from $2 \times 10^{19} \text{ W/cm}^2$ to $5 \times 10^{20} \text{ W/cm}^2$ and full width half maximum (FWHM) pulse durations $\tau_{FWHM}$ from $100 \text{ fs}$ to $500 \text{ fs}$. The FWHM laser spot size is set fixed for all simulations to $w_{FWHM} = 5 \text{ μm}$. This gives laser pulse energies $E_{laser}$ varying from $1.2 \text{ J}$ to $75 \text{ J}$, which are reachable with current high-power laser systems or future diode-pumped lasers at repetition rates of at least few shots per minute[15, 16]. A typical temporal laser profile can be seen in Fig. 1(b). Table I lists the laser parameters including pulse durations $\tau_{FWHM}$, peak intensities $I_0$ and pulse energy $E_{laser}$ for all simulations performed.

TABLE I. List of the laser parameters with fixed laser spot size $w_{FWHM} = 5\,\mu m$ for all simulations performed. The cells list the corresponding laser pulse energies. For $\tau_{FWHM} = 500$ fs a simulation at an intensity of $I_0 = 2.8 \times 10^{20}$ W/cm$^2$ was performed to better resolve the scaling of ion heating at long pulse durations, while the two simulations at $\tau_{FWHM} = 100$ fs were added to extrapolate the heating dynamics to ultra-short time scales. The parameters listed can be achieved by current high-power Ti:Sapphire laser systems or future diode-pumped high-power lasers with pulse repetition rates of more than a few shots per minute.

| $\tau_{FWHM}$ [fs] | $I_0$ [W/cm$^2$] | | | | |
|---|---|---|---|---|---|
| | $2\times10^{19}$ | $5\times10^{19}$ | $1\times10^{20}$ | $2.8\times10^{20}$ | $5\times10^{20}$ |
| 500 | 3 J | 7.5 J | 15 J | 42 J | 75 J |
| 400 | 2.4 J | 6 J | 12 J | - | 60 J |
| 300 | 1.8 J | 4.5 J | 9 J | - | 45 J |
| 200 | 1.2 J | 3 J | 6 J | - | 30 J |
| 100 | - | - | 3 J | - | 15 J |

At the laser wavelength of $\lambda_0 = 1.054\,\mu m$ chosen for all simulations the critical plasma density is $n_c = \dfrac{m_e \omega_0^2}{4\pi e^2} = 1.0 \times 10^{21}$ cm$^{-3}$ (CGS unit), with $m_e$, $e$, $\omega_0$ being the electron mass, charge and laser electric field angular frequency, respectively. Accordingly, ion number densities for deuteron, carbon and aluminium ions are set to realistic densities of $80n_c$, $40n_c$ and $60n_c$, corresponding to mass densities of the CD$_2$ layer of $1.1$ g/cm$^3$ and of the Al layer of $2.7$ g/cm$^3$. For the fully ionized target this yields electron number densities in the CD$_2$ layer and Al layer of $320n_c$ and $780n_c$, respectively.

All simulations start with an initially cold neutral plasma and initial ion charge states of $D^{+1}$, $C^{+1}$ and $Al^{+1}$ and include field ionization using the Landau-Lifshitz model[17, 18] and collisional ionization based on the Thomas-Fermi model [19].

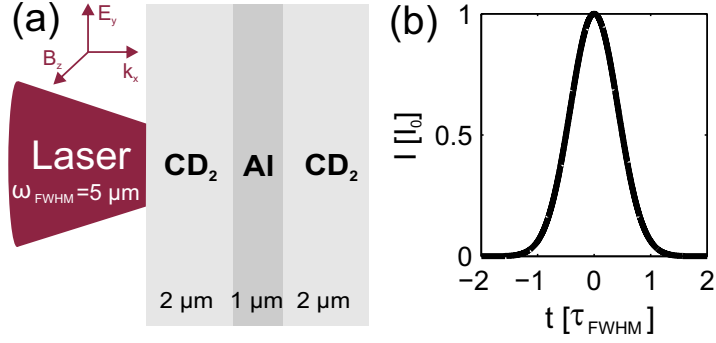

FIG.1. (a) Buried layer target structure as used in all simulations: The p-polarized laser pulse coming from the left impinges on the front surface of the target at normal incidence. (b) Temporal laser intensity profile. The time t = 0 corresponds to peak intensity on front surface of the target, which will be used as the reference time throughout the text.

## B. Numerical methods

The simulation box in our simulations consists of $N_x \times N_y = 9000 \times 4500$ cells. We employ absorbing boundary conditions and set the cell size to $\Delta x \times \Delta y = (\lambda_0 / 150) \times (\lambda_0 / 150)$ and the time step to $\Delta t = \Delta x / c = \Delta y / c$, with $c$ being the speed of light. This gives a spatial resolution of less than a quarter of the plasma wavelength $\lambda_{pe} / 4 \sim \lambda_0 / 112$. The number of macro electron particle per cell is 234 in the Al layer and 96 in the $CD_2$ layer when fully ionized, with each macro particle representing about one thousand real particles. We smooth macro-particle shapes by fourth-order splines and accordingly use fourth-order smoothing for the current deposition. With these parameters we achieve very good energy conservation over the complete simulation time, see Fig. 2.

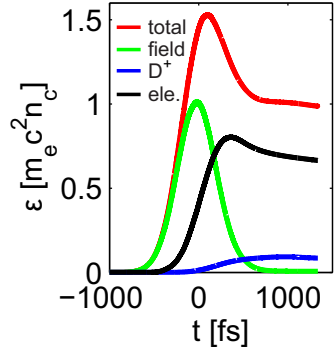

FIG.2. Energy density evolution in the case of laser intensity $I_0 = 2\times10^{19}$ W/cm$^2$ and pulse duration $\tau_{FWHM} = 500$ fs. Around 1ps the field energy almost drops to zero and is fully converted into electron and ion kinetic energy. No numerical heating is observed as the particle kinetic energies stay constant after 1ps. The small decrease in total energy after 1ps is mainly due to particles leaving the simulation box.

## C. Modelling radiative energy loss

We have estimated the effect of radiation loss by FLYCHK[20] simulations and find that it does not affect the electron temperature significantly and can thus be ignored in simulations. Fig. 3(a) shows the radiation power containing bound-bound radiation $Pr_{bb}$, free-bound radiation $Pr_{fb}$ (recombination), free-free radiation $Pr_{ff}$ (Bremsstralung radiation) and total radiation $Pr_{tot} = Pr_{bb} + Pr_{fb} + Pr_{ff}$ as a function of the electron temperature in Al layer. As an example, for the simulation with parameters $I_0 = 10^{20}$ W/cm$^2$ and $\tau_{FWHM} = 500$ fs the maximum bulk electron temperature reaches 39 keV and the total radiated power $Pr_{tot}$ on a time scale of 2 ps yields an estimated energy loss of ~ 2 mJ which is ~ 0.01% of the 15 J laser energy. The resulting radiation power loss per electron of around 300 eV/ps/electron is less than 10% of the total bulk electron temperature for all the simulations. As Fig. 3(b) suggests, the target is optically thin for all photon wavelengths not resolved in the particle-in-cell simulations. We thus find that all radiative loss is well accounted for in our simulations and that the target does not reach a radiative equilibrium state.

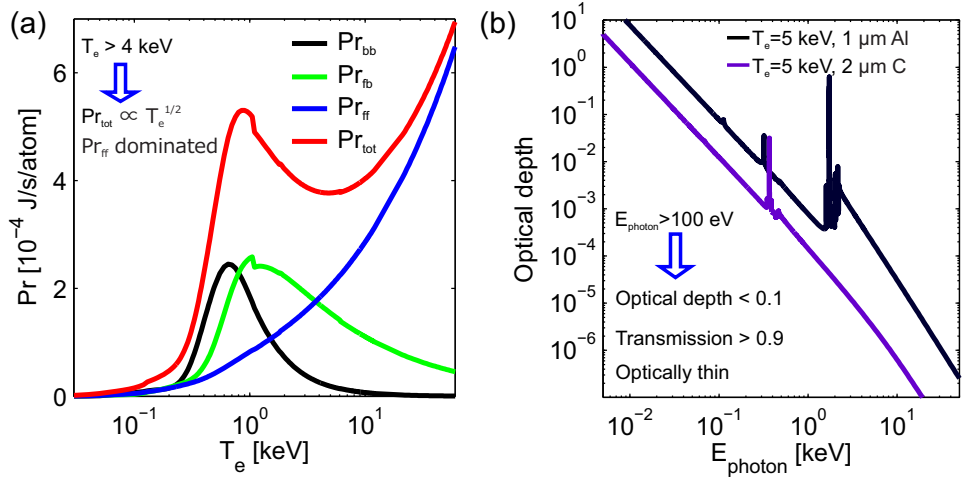

FIG.3. (a) Radiative power versus electron temperature in the Al layer: At low temperatures, radiation is dominated by bound-bound transitions which rise with electron temperature. As more and more Al are ionized radiation becomes dominated by Bremsstrahlung which is approximately proportional to $T_e^{1/2}$. (b) Optical depth for 1 μm aluminium and 2 μm Carbon versus photon energy at 5 keV electron temperature, the minimum saturated bulk electron temperature observed in simulations: For photon energies above 100 eV the transmission becomes greater than 90% and the target is optically thin. Electromagnetic radiation below 180 eV is resolved on the simulation grid.

## III. OVERVIEW OF TARGET IONIZATION AND HEATING

We start by discussing a specific simulation with laser parameters $I_0 = 10^{20} \text{ W/cm}^2$ and $\tau_{FWHM} = 500 \text{ fs}$ illustrating the heating dynamics inside the buried layer target before focusing on the scaling of target heating with the laser parameters. A schematic picture of hot electron acceleration, bulk ionization, return current generation, bulk electron heating and electron pressure jump formation at the interface of $CD_2$ layer and Al layer is presented in Fig. 4(a)~(e), while Fig.4 (f)~(j) show the corresponding processes extracted from simulations.

As can be inferred from the distribution of electron longitudinal phase space density in Fig. 4(f), bunches of hot electrons with relativistic velocities are driven into the target at a rate of twice the laser frequency by the laser magnetic force as soon as the laser intensity on target front reaches relativistic values. This current of hot electrons ionizes both the Al and $CD_2$ layer over a time of several hundred femtoseconds by collisional ionization, as can be seen from the corresponding temporal evolution of the free electron density inside the target in Fig. 4(g), which reaches its maximum well before the laser pulse peak intensity reaches the target. The same hot electron current drives a resistive electrostatic field[21] that in turn drives a cold bulk electron return current balancing the hot electron current[22, 23]. In Fig. 4(h), the transverse velocity distribution of electrons in the z-direction is compared to the distribution in the longitudinal x-direction in the velocity range $-0.2c < v_x < 0.2c$. While the transverse electron dynamics are governed by random thermal motion, the longitudinal dynamics include the hot electron current and cold bulk electron current. The longitudinal velocity distribution shows a distinct asymmetry indicating an increase in slow electrons moving backwards to the target front side (in negative x-direction) as expected for a return current counteracting the high-energy forward moving electrons. Note that Fig. 4(h) is extracted from a simulation with pulse duration $\tau_{FWHM} = 100$ fs at the time $-22.7$ fs, where the asymmetry is more pronounced than for longer pulses.

The collision rate[24] $v_e = \frac{4\sqrt{2\pi} \, n_e e^4 \ln\Lambda}{3 \, m_e^{1/2} (k_B T_e)^{3/2}} > 25 \text{ ps}^{-1}$ and mean free path $l_e = \frac{3(k_B T_e)^2}{4\sqrt{2\pi} \, n_e e^4 \ln\Lambda} < 5 \, \mu\text{m}$, $\ln\Lambda$ being the Coulomb logarithm, for electrons within the velocity range $-0.2c < v_x < 0.2c$ indicate that it is mainly electrons in that range which contribute to bulk electron heating. Fig. 4(i) shows the temporal evolution of the bulk electron temperature $T_{bulk}$ increasing gradually during the intra-pulse phase, saturating at 39 keV in the tail of laser pulse. All temperatures are calculated non-relativistically by $k_B T = m v_{th}^2 / 2$ where $v_{th}$ is the width of the particle velocity distribution in z direction, which is orthogonal to both the direction of laser propagation and laser polarization.

The formation of an electron pressure ($P_e = n_e k_B T_e$) gradient at the interface of the $CD_2$ layer and Al layer can be seen in Fig. 4(j), which gives the average electron pressure within $2\,\mu m$ around the laser focus center at $-184\,fs$. This pressure difference mainly comes from the electron density difference between the two layers, as the bulk electron heating heats both layers to very similar temperatures, and drives an expansion wave propagating into the $CD_2$ layer.

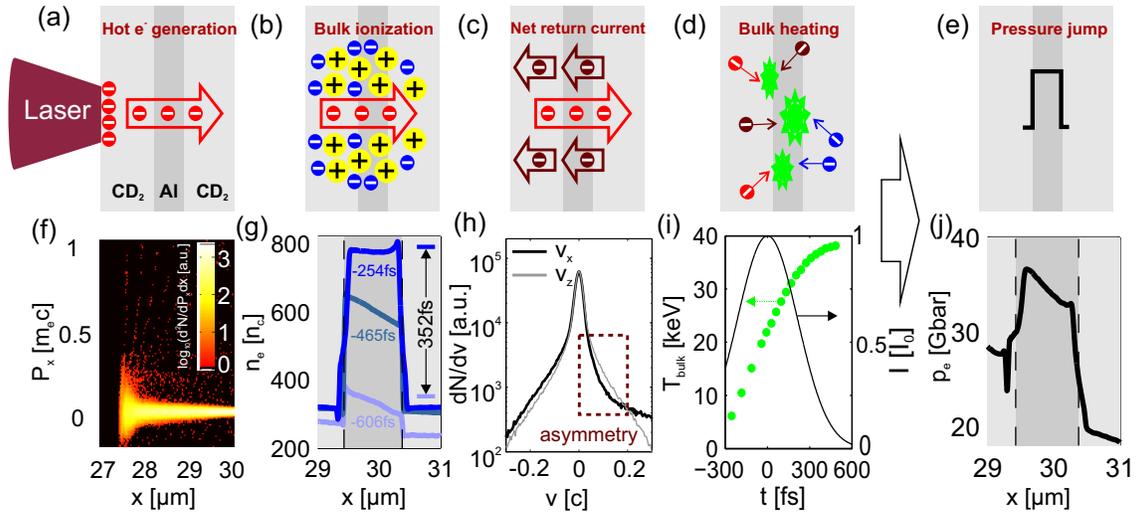

FIG.4. Schematic illustration of (a) hot electron generation, (b) bulk ionization, (c) return current generation, (d) bulk electron heating and (e) electron pressure jump formation; (f) Electron longitudinal phase space density at $-606\,fs$; (g) Free electron density evolution at $-606\,fs$, $-465\,fs$ and $-254\,fs$; (h) Electron velocity distribution $dN/dv$ in x and z direction at $-22.7\,fs$ (laser parameters $I_0 = 10^{20}\,W/cm^2$ and $\tau_{FWHM} = 100\,fs$); (i) Temporal evolution of bulk electron temperature averaged over the whole target depth; (j) Electron pressure along the laser axis from a $2\,\mu m$ average over the y direction ($14\,\mu m \leq y \leq 16\,\mu m$) at $-184\,fs$. For details please see discussion in the text.

## A.  Pressure gradient formation and deuteron layer expansion

In order to illustrate the expansion of the buried layer in the target due to the pressure gradient, Fig. 5(a) shows the density of deuterons, aluminium and electrons along the laser axis from a 2 µm average over the range from 14 µm to 16 µm in y-direction at 343.9 fs, while Fig. 5(c) shows the corresponding two dimensional distribution of deuteron density, aluminium density at the same time. During the expansion phase, a distinct interface between the Al and $CD_2$ layer is observable, marked by the dashed line in Fig. 5(a) and clearly visible in Fig. 5(c). In the following we will focus on the dynamics of the expansion in the second $CD_2$ layer facing away from the laser and note that most of our findings also apply to the first layer.

No mixing of the aluminium and deuteron ions across this interface is observed. The Al layer acts like a piston pushing the $CD_2$ layer forward with the material velocity $v_m = 1.68 \times 10^{-3} c$. During the expansion of the Al layer, a clear drop in both the electron and deuteron density appears, marked by the black arrow in Fig. 5(a), which we from now on call the "expansion front". The region between the Al/$CD_2$ interface and the expansion front will be called "expansion region". The density of electrons in the expansion region is compressed by almost a factor of $\alpha \approx 1.5$. A similar increase can be seen in the deuteron density. The expansion front at all times coincides with a spike in the longitudinal electrostatic field shown in Fig. 5(b). This field accelerates the deuteron ions forward. It spreads over a width of $\sim 0.05\,\mu m$ which is close to the deuteron ion mean free path $l_i \sim 0.07\,\mu m$. The velocity of the expansion front $v_{exp} = 4.7 \times 10^{-3} c$, extracted from the linear fit in Fig. 5(d) to the various positions of the expansion front at different times, agrees with the value derived from mass flux conservation[25] $(\alpha - 1)v_{exp} = \alpha v_m$ and is always close to the ion acoustic velocity[24] $C_s = (Z k_B T_{bulk}/M)^{1/2}$, with $Z/M$ being the charge-to-mass ratio of the deuteron and carbon ions respectively. In all simulations presented here, the Mach number $\Gamma = v_{exp}/C_s$ is very close to unity, indicating that, in contrast to previous works[26], the expansion wave is not a shock wave and the enhanced deuteron heating in the expansion region does not require a shock to develop.

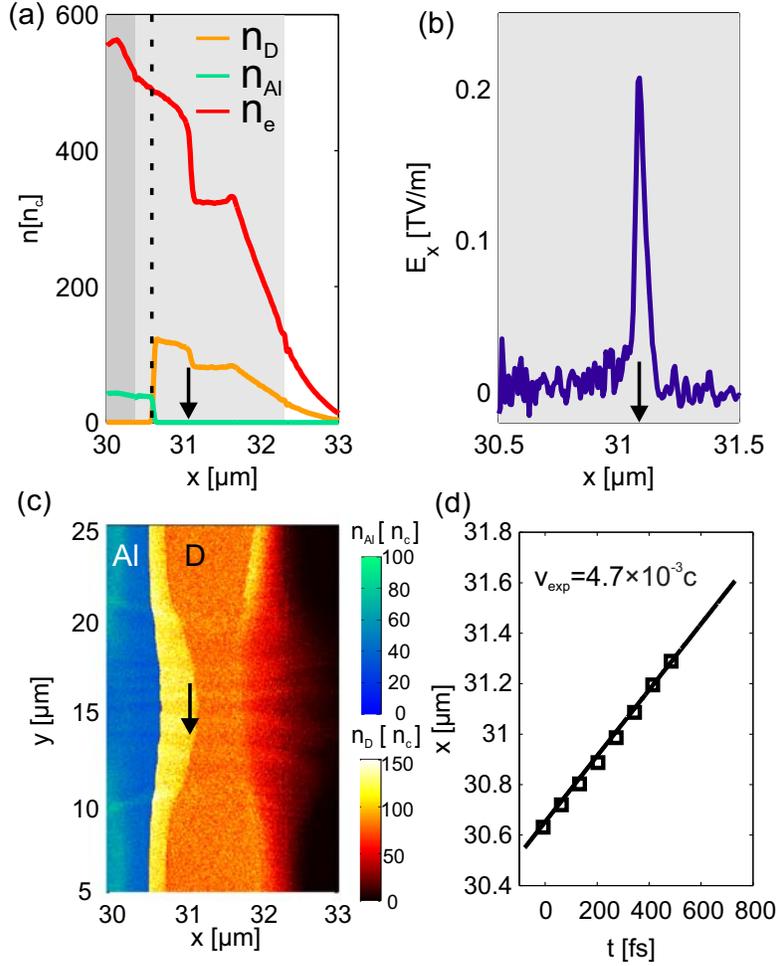

FIG.5. (a) Density of deuteron ions, aluminium ions and electrons along the laser axis from a 2 μm over the range from 14 μm to 16 μm in y-direction at t= 343.9 fs , the dashed vertical line marks the interface between the Al layer and the $CD_2$ layer at the same time; (b) Electrostatic field along the laser axis at the same time as in (a), averaging over the y direction in a similar fashion; (c) Corresponding colour-coded density distribution of deuteron and aluminium ions, clearly showing the interface and the expansion front marked by the black arrow and the rarefaction wave coming from the right side of the target and counterpropagating to the expansion. (d) Temporal evolution of the expansion front.

## B. Expansion-driven deuteron heating

As pointed out in the previous section, deuteron heating does not require the formation of a shock front during the expansion of the deuterons driven by the expanding Al layer. Rather, we find that the expansion of the Al layer simply provides the energy source feeding the deuteron heating.

Following the phase space plot of the deuteron density as indicated in Fig. 6(a) we find that both the deuteron ions in the expansion region and in the bulk show a Maxwellian velocity distribution, see Fig. 6(b). Both the phase space plot and the velocity distribution clearly show a higher deuteron temperature in the expansion region. Although both the electron density and deuteron density are increased in this region compared to the target bulk, this increase is insufficient to explain the increased heating with binary collisions using the Spitzer collision rate[24].

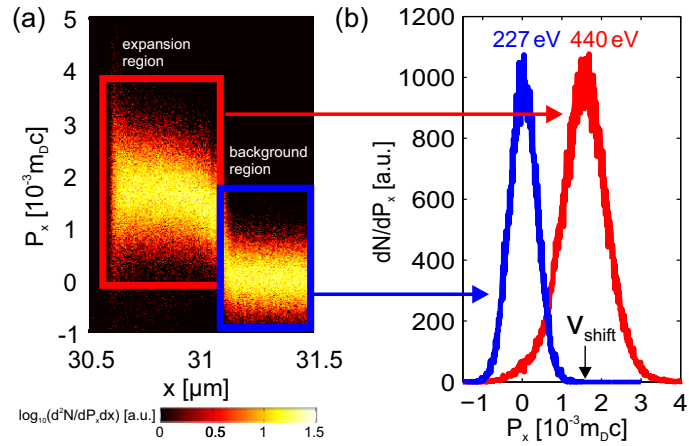

FIG.6. (a) Deuteron longitudinal phase space density at the same time as in Fig. 5, the red and blue rectangles enclose the expansion region and background region, respectively; (b) The corresponding deuteron momentum distribution in the two regions. Both distributions are Maxwellian with the corresponding temperatures in eV cited. As can be clearly seen, the expansion region shows a beam-like drift, with the black arrow indicating the shift velocity relative to the resting background ions.

The expansion velocity is always higher than the material velocity. As a consequence, during the expansion the expansion region itself grows in size along the direction of expansion, with more and more deuteron ions from the bulk entering the expansion region. On closer inspection, one finds a shift $v_{shift}$ in deuteron velocity between the bulk and the expansion region which comes from the acceleration by the longitudinal electrostatic field at the expansion front shown in Fig. 5(b), around which the deuteron velocity distribution in the latter region is centred. During the continuous injection of ions, energy is transferred from this directed, beam-like motion of the expansion region into thermal motion of ions in this region, as shown in the following. To begin with, Fig. 7 shows the temporal evolution of deuteron temperature in the expansion region for varying laser pulse duration and intensity, with each time series truncated at the point the expansion region merges with the rarefaction wave from the backside of the target. The deuteron temperature increase is found to be faster for higher laser intensities and fixed pulse duration, coinciding with an earlier start of the expansion and a higher expansion velocity.

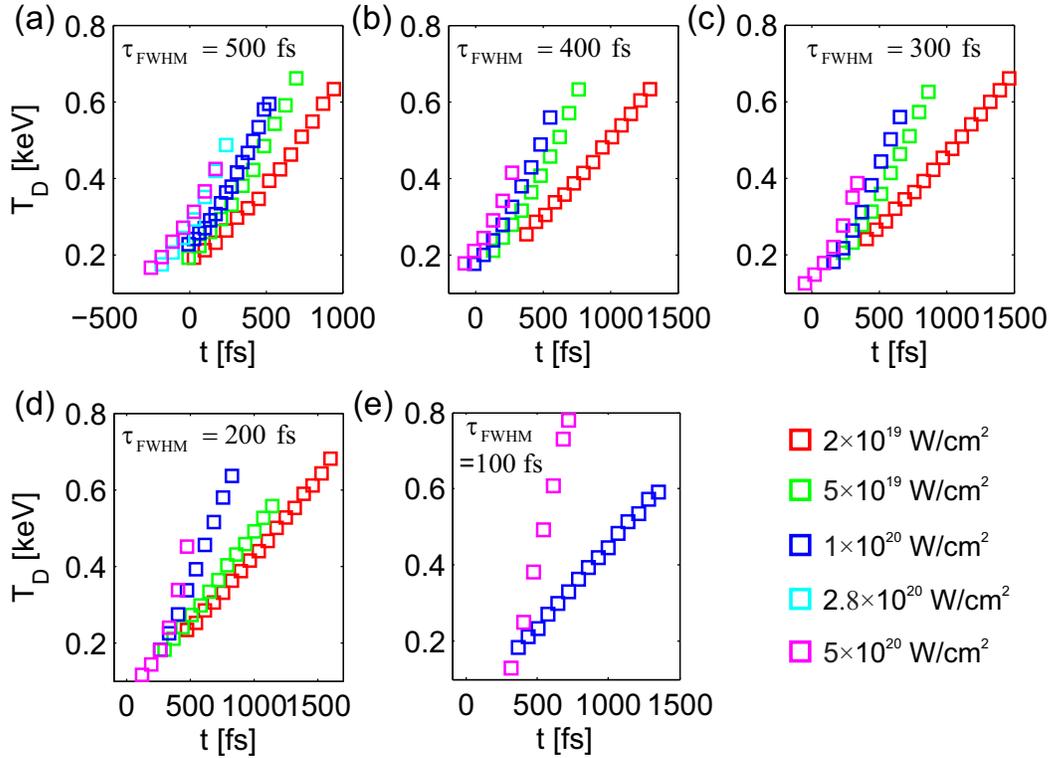

FIG.7. Scaling of deuteron temperature in the expansion region with time for various laser intensities and laser pulse durations. When the laser pulse is still transferring energy to the target, the deuteron energy increases

following a parabolic scaling. As soon as the laser no longer impinges on the target, deuteron temperature increases linearly. Each time series is truncated at the point the expansion region merges with the rarefaction wave from the backside of the target.

Fig. 8 shows the deuteron kinetic beam energy $E_{shift} = m_D v_{shift}^2 / 2$ corresponding to the directed, beam-like motion of deuterons in the expansion region for the same set of laser parameters and the same time period. If we compare the trend of temperature and kinetic beam energy of the deuterons, one finds a linear increase in kinetic beam energy and a quadratic increase in temperature during the time the laser transfers energy to the target. After the laser pulse has delivered all its energy to the target, the kinetic beam energy remains almost constant with only a small decrease over time while the temperature increases linearly.

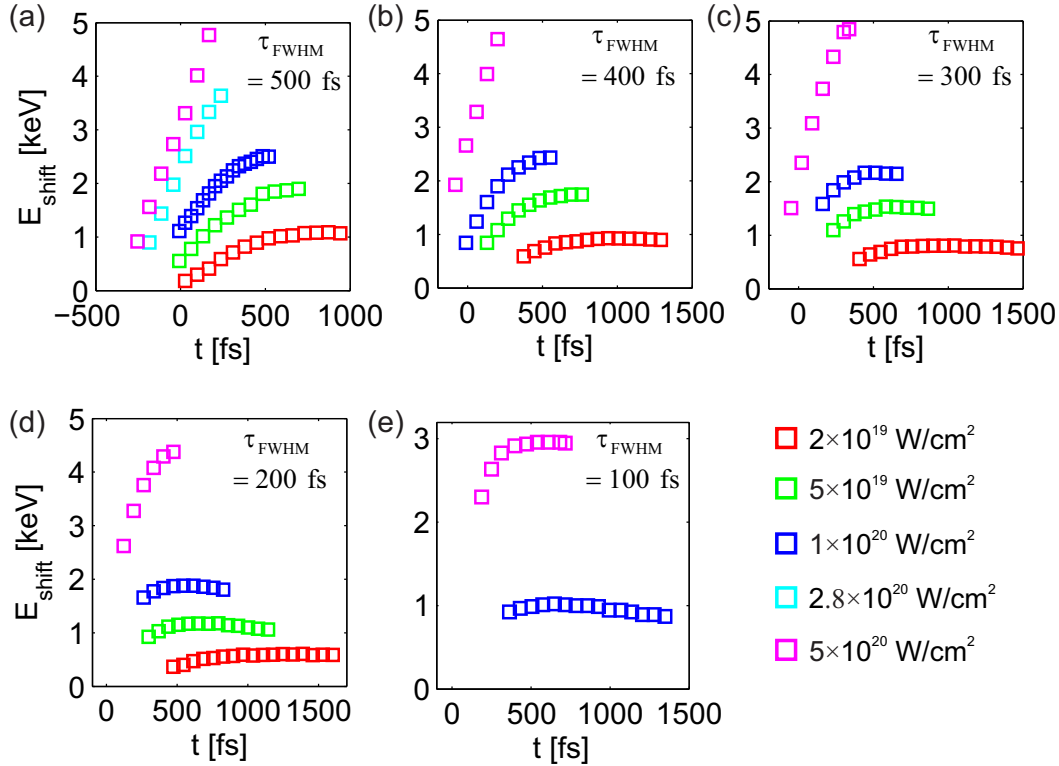

FIG.8. Scaling of deuteron kinetic beam energy $E_{shift} = m_D v_{shift}^2/2$ in the expansion region with time for various laser intensities and laser pulse durations: When compared to Fig. 7, one clearly sees that the kinetic beam energy rises linearly when the laser still transfers energy to the target and remains almost constant after the laser pulse has deposited its energy. Each time series is truncated at the point the expansion region merges with the rarefaction wave from the backside of the target.

When comparing this temperature increase to the increase due to electron-ion heating by collisions, we find the former always to be much stronger than the latter, making it possible to ignore electron-ion heating for the following discussion. Consequently, we postulate a time-dependent rate $R(t)$ of energy transfer between deuteron kinetic beam energy and temperature in the expansion region

$$\frac{dT_D(t)}{dt} = R(t)E_{shift}(t) \qquad (1)$$

for which integration yields

$$T_D(t) = \int R(t)E_{shift}(t)dt + const. \qquad (2)$$

Following the trends in Fig. 7 and Fig. 8, we find the energy transfer rate to be almost a constant $R$ independent of time, while the offset to the deuteron temperature is simply given by the temperature of the bulk deuteron ions due to electron-ion heating at the time the expansion starts.

This constant rate $R$ is extracted from the data using Eq. (2) and plotted in Fig. 9 with the error bars including both the small variation of $R$ over time for each fit and the fitting error. Within these error bars, we find for most pulse durations a steady decrease in $R$ with laser intensity. One has to keep in mind that this decrease is counteracted by the increase in kinetic beam energy with laser intensity depicted in Fig. 8, resulting in a net increase in deuteron temperature with laser intensity. For the case of the shortest pulse duration of 100 fs this trend is clearly broken, showing a strong increase in $R$ at high laser intensities.

We have compared the energy transfer rate to the collision rate between ions in the bulk and the expansion region, and to the collision rate between ions in the expansion region only. Both rates are typically on the same order of magnitude as the rate $R$, meaning that deuteron and carbon ions can be considered to be in thermal equilibrium. The collisional rates however depend strongly on the ion kinetic beam energy and the ion temperature, which makes the collisional energy transfer process complex and thus difficult to model. As we have ruled out any numeric heating, we attribute the deuteron heating in the expansion region to a combination of several collisional processes and collective plasma effects that we will investigate in a follow-up analysis which would be beyond the scope of this publication. Although the heating dynamics in the expansion region are complex, it is remarkable that we can describe them by a simple effective energy transfer rate.

We nevertheless find that for long pulses the clear decrease of $R$ with laser intensity and the accompanying increase in $v_{shift}$ hints to a random energy transfer process between ions in the expansion region and the bulk region with a cross section decreasing with velocity, similar as one would expect for collisional energy transfer.

Our data suggests a change in energy transfer for the case of high laser intensities at ultra-short pulse durations. This change is significant but requires further investigation into the heating mechanism at work.

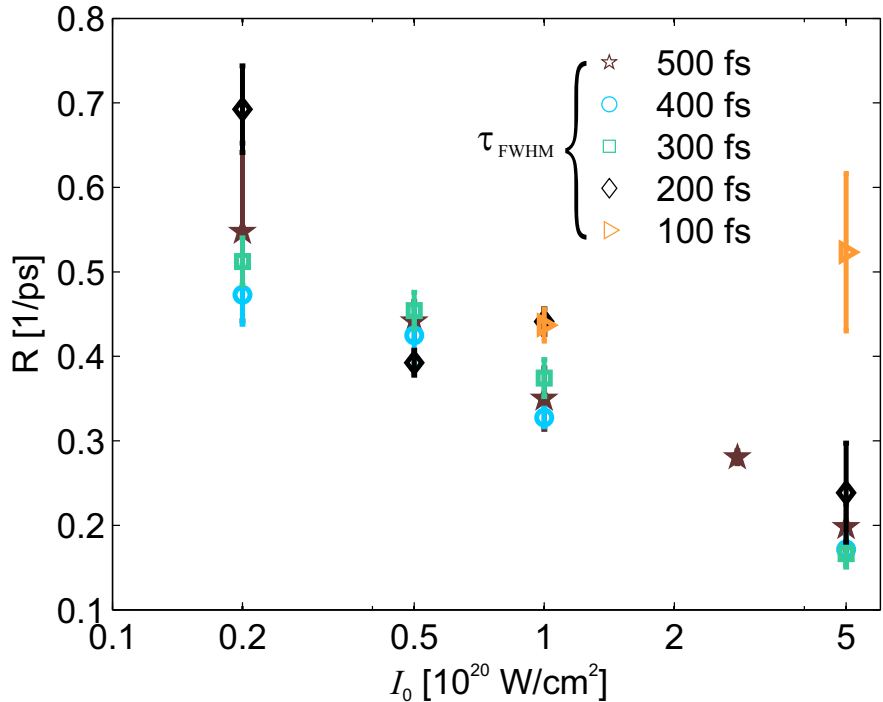

FIG.9. Energy transfer rate $R$ from kinetic beam energy to temperature of the deuteron ions in the expansion region as a function of laser intensity in different pulse duration cases. With increasing laser intensity the transfer rate decreases, while the kinetic beam energy increases, resulting in a net increase in deuteron temperature with laser intensity. The trend of decreasing rate $R$ with laser intensity is broken for the very short pulse duration of 100 fs, where we see an increase in $R$ for very high laser intensities, possibly indicating the onset of a new heating mechanism.

## IV. CONCLUSION AND OUTLOOK

Our investigation of ion heating in buried layer targets irradiated with high-intensity, short pulse lasers has shown that strong ion heating to several hundred eV is in reach with high-power, short-pulse laser systems that today can deliver few shots per minute. This opens up the possibility to study the ultra-short dynamics of ion heating with dedicated campaigns of several hundred shots, entering the realm of high statistics under reproducible experimental conditions.

As such laser systems become available at an increasing number of facilities, experiments no longer have to focus on the heating dynamics in bulk material in the several picosecond to nanosecond range, but can instead probe the non-equilibrium, feature-rich processes happening on the sub-picosecond time scale.

Studying buried layer targets we have shown that the electron heating processes during the laser interaction with the target can be decoupled from ion heating in the bulk by storing part of the energy as a pressure difference between the different target layers, which allows for the expansion of the inner target layer and the subsequent heating processes on the time scale of the ion acoustic velocity. This time scale could potentially be probed by existing, ultra-short X-ray lasers, enabling detailed studies of the temporal evolution of the plasma dynamics[27].

We have found that we can correlate the beam-like expansion of the deuteron ions that is driven by the aluminium layer expansion to the heating of deuteron ions in the expansion region. With the introduction of a simple, constant energy transfer rate from kinetic beam energy to thermal energy, we are able to derive the temporal evolution of the temperature from the temporal evolution of the kinetic beam energy of the deuteron ions.

Moreover, we find indications of a change from this simple energy transfer model when going to ultra-short pulse durations at high intensities. This not yet understood sudden change will be subject of future studies that are beyond the scope of the analysis presented here. The richness of ion and electron dynamics in buried layer targets irradiated with high-intensity pulses on femtosecond could prove a new and exciting field of non-equilibrium plasma physics at solid densities which can be investigated with compact, high-intensity laser sources with repetition rates of a few shots per minute.

## ACKNOWLEDGMENTS

This work was partially supported by the joint research project onCOOPtics under Grant No. 03ZIK445, the Natural Science Foundation of China under Grant No. 11174303, and the National Basic Research Program of China (973 Program) under Grant Nos. 2008CB717806 and 2011CB808104. The authors would like to thank the HZDR HPC group for their support.